\documentclass[manuscript]{aastex}
\usepackage{rotating,graphicx}

\shorttitle{Longitudinally Asymmetric Distribution of Solar Energetic Particles}

\shortauthors{He \& Wan}

\begin{document}

\title{Numerical Study of the Longitudinally Asymmetric Distribution of Solar Energetic Particles in the Heliosphere\\}

\author{H.-Q. He\altaffilmark{1,2} and W. Wan\altaffilmark{1}}

\altaffiltext{1}{Key Laboratory of Earth and Planetary Physics,
Institute of Geology and Geophysics, Chinese Academy of Sciences,
Beijing 100029, China; hqhe@mail.iggcas.ac.cn,
wanw@mail.iggcas.ac.cn}

\altaffiltext{2}{CAS Key Laboratory of Geospace Environment,
Department of Geophysics and Planetary Sciences, University of
Science and Technology of China, Hefei, Anhui 230026, China}

\begin{abstract}
Solar energetic particles (SEPs) affect the solar-terrestrial space
environment and become a very important aspect in space weather
research. In this work, we numerically investigate the transport
processes of SEPs in three-dimensional interplanetary magnetic
field, with an emphasis on the longitudinal distribution of SEPs in
the heliosphere. We confirm our previous finding that there exists
an east-west longitudinal asymmetry in the SEP intensities, i.e.,
with the same longitude separations between the solar source centers
and the magnetic footpoint of the observer, the fluxes of SEP events
originating from solar sources located on the eastern side of the
nominal magnetic footpoint of the observer are systematically larger
than those of the SEP events originating from sources located on the
western side. We discuss the formation mechanism of this phenomenon,
and conclude that the longitudinally asymmetric distribution of SEPs
results from the east-west azimuthal asymmetry in the topology of
the heliospheric magnetic field as well as the effects of
perpendicular diffusion on the transport of SEPs in the heliosphere.
Our results will be valuable to understanding Sun-Earth relations
and useful for space weather forecasting.
\end{abstract}


\clearpage

\section{Introduction}
Solar energetic particles (SEPs), which are charged energetic
particles occasionally emitted by the Sun, risk the health of
astronauts working in space and damage electronic components on
satellites, so they have become a very important aspect affecting
solar-terrestrial space environment and space weather.
Theoretically, SEPs observed in the interplanetary magnetic field
provide fundamental information regarding the acceleration
mechanisms and transport processes of charged energetic particles.
Therefore, the subject has become a focus of astrophysics, space
physics, and plasma physics.

Through several decades of investigations in the community with
spacecraft observations and theoretical modeling, significant
progresses have been achieved toward a better understanding of the
transport processes of SEPs. As a pioneering work,
\citet{Parker1965} provided a diffusion equation to investigate the
modulation of galactic cosmic rays and the transport of SEPs. It is
very difficult to analytically solve the multidimensional transport
equation; consequently, numerical calculations are usually adopted
\citep[e.g.,][]{Ng1979,Ruffolo1991,Zhang2009,He2011a,He2011b,Giacalone2012}.
\citet{Jokipii1966} proposed the well-known and classical
quasi-linear theory (QLT) to describe the diffusion of charged
particles in a turbulent magnetic field. The theoretical
investigations and observational analyses based on SEP events
measured by the Helios spacecraft suggested some nonlinear effects
near $90^{\circ}$ pitch angle in the particle pitch-angle
distributions
\citep[e.g.,][]{Hasselmann1968,Hasselmann1970,Beeck1986,Beeck1987}.
\citet{Schlickeiser2002} laid complete physical foundations and
presented profound mathematical techniques for cosmic ray transport
research. \citet{Matthaeus2003} provided a nonlinear guiding center
(NLGC) theory for charged particles' perpendicular diffusion to
account for the numerical simulation results of test particles.
\citet{Shalchi2004} presented analytical forms for the results from
the NLGC theory. \citet{Zank2000} and \citet{Li2003} provided a
dynamical time-dependent model of charged particle acceleration at a
propagating and evolving interplanetary shock.

Recently, \citet{Zhang2009} presented a model calculation of SEP
propagation in a three-dimensional interplanetary magnetic field
with the effect of magnetic turbulence. \citet{He2011b} investigated
the effects of particle source characteristics on SEP observations
at 1 AU and found that the perpendicular diffusion has a very
important influence on the propagation of SEPs in the heliosphere,
particularly when a spacecraft is not directly connected to the
acceleration regions either on the Sun or near the CME-driven shocks
by the interplanetary magnetic field lines; in such cases, the
earliest arriving particles can be seen propagating toward the Sun,
having scattered backward at large distances. \citet{He2012a}
presented a direct method to quickly and explicitly determine the
spatially dependent parallel and radial mean free paths of SEPs with
adiabatic focusing. Furthermore, \citet{He2012b} provided a direct
approach to explicitly investigate the spatially dependent
perpendicular mean free path of SEPs in a turbulent and spatially
varying magnetic field. They reported that for physical conditions
representative of the solar wind and the magnetic turbulence, the
ratio $\lambda_{\perp}/\lambda_{\parallel}$ of the perpendicular to
the parallel mean free path remains in the range $0.01-0.20$;
however, when the turbulence strength $\delta B/B$ is sufficiently
large, the ratio $\lambda_{\perp}/\lambda_{\parallel}$ would
approach or exceed unity.

Observationally, \citet{Cane1988} investigated time profiles of 235
intense proton events to document the longitude-dependent profiles.
\citet{Reames1999} presented some observational evidences of
longitudinal effects on the transport processes of SEPs. They found
that the interplanetary spiral magnetic field causes an asymmetry in
the intensity time profiles of SEP events originating from eastern
and western longitudes on the solar disk. As we know, SEPs are
usually produced by solar flares or/and related to coronal mass
ejections (CMEs) \citep[e.g.,][]{Reames1999}. Recently,
\citet{Dresing2012} investigated in detail the large longitudinal
spread of SEPs during the 17 January 2010 solar event and suggested
with observational evidence that the wide longitudinal spread of
particles is mainly due to perpendicular diffusion in the
interplanetary medium.

In \citet{He2011b}, the authors found that with the same separation
in heliographic longitude between the magnetic footpoint of the
observer and the solar sources, the SEPs associated with the sources
located east are detected earlier with larger fluxes than those
associated with the sources located west. This interesting
phenomenon was called the ``east-west azimuthal asymmetry of SEPs"
by \citet{He2011b}. The simulation results in \citet{Giacalone2012}
also showed similar asymmetric features of SEP distribution. In this
paper, we focus on the so-called east-west azimuthal asymmetry of
SEPs in view of numerical investigations. There are several previous
works studying the longitudinal dependence of SEP intensities
\citep[e.g.,][]{Cane1988,Reames1999}. Basically, however, these
previous studies only analyzed the effects of heliographic longitude
of particle sources on the intensity profiles of SEP events. In our
simulations, we mainly investigate the relatively longitudinal
distribution of SEP intensities in the heliosphere, and find the
longitudinally asymmetric distribution phenomenon. With the same
longitude separations between the solar source centers and the
magnetic field line footpoint of the observer, the fluxes of SEP
events originating from solar sources located on the eastern side of
the nominal magnetic footpoint of observer are systematically larger
than those of the SEP events originating from sources located on the
western side. We conclude that the longitudinally asymmetric
distribution of SEPs results from the east-west azimuthal asymmetry
in the geometry of the heliospheric magnetic field as well as the
effects of perpendicular diffusion on the transport processes of
SEPs in the heliosphere. Our results will be valuable for
understanding solar-terrestrial relations and useful in space
weather forecasting. Some investigation results related with the
statistical and numerical studies on the longitudinally asymmetric
distribution of SEPs were reported in the 33rd International Cosmic
Ray Conference \citep{He2013a}.

This paper is organized as follows. Firstly, in Section 2, we
present a five-dimensional focussed transport model containing most
of the usually used transport mechanisms. The simulation method for
numerically solving the focussed transport equation is also provided
in this section. In Section 3, the simulation results of the
relatively longitudinal distribution of SEPs will be presented. We
present some implications of our simulation results in Section 4 and
a discussion on the formation mechanism in Section 5. A summary of
our results will be provided in Section 6.

\section{Numerical Model and Method}
In our model, the five-dimensional focussed transport equation that
governs the gyrophase-averaged distribution function
$f(\textbf{x},\mu,p,t)$ of SEPs can be written as
\citep[e.g.,][]{Schlickeiser2002,Zhang2009,He2011b,Qin2011}
\begin{eqnarray}
{}&&\frac{\partial f}{\partial t}+\mu v\frac{\partial f}{\partial
z}+{\bf V}^{sw}\cdot\nabla f+\frac{dp}{dt}\frac{\partial f}{\partial
p}+\frac{d\mu}{dt}\frac{\partial f}{\partial \mu}  \nonumber\\
{}&&-\frac{\partial}{\partial\mu}\left(D_{\mu\mu}\frac{\partial
f}{\partial \mu}\right)-\frac{\partial}{\partial
x}\bigg(\kappa_{xx}\frac{\partial f}{\partial x}\bigg)
-\frac{\partial}{\partial y}\left(\kappa_{yy}\frac{\partial
f}{\partial y}\right)=Q({\bf x},p,t),  \label{transport-equation}
\end{eqnarray}
where $\textbf{x}$ is particle's position, $z$ is the coordinate
along the magnetic field spiral, $p$ is particle's momentum, $\mu$
is the pitch-angle cosine of particle, $t$ is time, $v$ is the
particle velocity, $\textbf{V}^{sw}$ is the solar wind speed, and
$Q$ is the source term. In the transport model, we directly input a
source of accelerated particles as a product of either a solar flare
or the CME-driven shock. We focus on the high-energy SEPs
accelerated and released into the system near the Sun. The particle
injection can be seen as a point source in the radial direction. For
observations far out near 1 AU or larger radial distances, this
assumption is particularly valid and is commonly used in the study
of SEP transport \citep[e.g.,][]{Zhang2009}. The particle momentum
change term $dp/dt$ in Equation (\ref{transport-equation}) due to
the adiabatic cooling effect can be written as \citep{Skilling1971}
\begin{equation}
\frac{dp}{dt}=-p\left[\frac{1-\mu^2}{2}\left(\frac{\partial
V^{sw}_x}{\partial x}+\frac{\partial V^{sw}_y}{\partial
y}\right)+\mu^2\frac{\partial V^{sw}_z}{\partial z}\right].
\label{adiabatic-cooling}
\end{equation}
In addition, $d\mu/dt$, which includes the magnetic focusing effect
and the divergence of the solar wind flows effect, can be expressed
as \citep[e.g.,][]{Roelof1969,Isenberg1997,Kota1997}
\begin{eqnarray}
\frac{d\mu}{dt}&=&\frac{1-\mu^2}{2}\left[-\frac{v}{B}\frac{\partial
B}{\partial z}+\mu \left(\frac{\partial V^{sw}_x}{\partial
x}+\frac{\partial V^{sw}_y}{\partial y}-2\frac{\partial
V^{sw}_z}{\partial z}\right)\right]     \nonumber\\
{}&=&\frac{1-\mu^2}{2}\left[\frac{v}{L}+\mu\left(\frac{\partial
V^{sw}_x}{\partial x}+\frac{\partial V^{sw}_y}{\partial
y}-2\frac{\partial V^{sw}_z}{\partial z}\right)\right],
\label{eq:magnetic-focussing}
\end{eqnarray}
where $B$ is the background interplanetary magnetic field with
direction $\textbf{z}$, and the magnetic focusing length $L$ is
defined as $L=(\textbf{z}\cdot\bigtriangledown \ln B)^{-1}$.
Therefore, the adiabatic cooling effect and the divergence of the
solar wind flows effect generated by the solar wind velocity
components, expressed as Equations (\ref{adiabatic-cooling}) and
(\ref{eq:magnetic-focussing}), respectively, have been included in
our model. The interplanetary magnetic field in the inner
heliosphere is too strong and the particle energies in typical SEP
events are not high enough, so in this work we neglect the particle
drift in the nonuniform magnetic field \citep{Zhang2009}.

Under the diffusion approximation for a nearly isotropic pitch-angle
distribution, the parallel mean free path $\lambda_{\parallel}$ can
be written as \citep{Jokipii1966,Hasselmann1968,Earl1974,He2014}
\begin{equation}
\lambda_{\parallel}=\frac{3v}{8}\int_{-1}^{+1}\frac{(1-\mu^{2})^{2}}{D_{\mu\mu}}d\mu.
\label{parallel-path}
\end{equation}
Accordingly, the radial mean free path can be defined as
\begin{equation}
\lambda_{r}=\lambda_{\parallel}\cos^{2}\psi, \label{radial-path}
\end{equation}
where $\psi$ is the angle between the local magnetic field direction
and the radial direction. In Equation (\ref{radial-path}),
$\cos^{2}\psi$ can be written as \citep{Ng1971,He2012a}
\begin{equation}
\cos^{2}\psi=(V^{sw})^2/\left((V^{sw})^2+\Omega^2r^2\sin^2\theta\right),
\label{cos2psi}
\end{equation}
where $V^{sw}$ is the solar wind speed, $\Omega$ is the angular
rotation velocity of the Sun, and $r$ and $\theta$ are the
coordinates of the heliocentric spherical coordinate system
$(r,\theta,\phi)$, namely, $r$ is the heliocentric radial distance,
and $\theta$ is the colatitude, which is measured from the rotation
axis of the Sun.

We use a form of the pitch-angle diffusion coefficient as
\citep[e.g.,][]{Beeck1986,He2011b}
\begin{equation}
D_{\mu\mu}^{r}=D_{\mu\mu}/\cos^{2}\psi=D_{0}vR^{-1/3}\left(|\mu|^{q-1}+h\right)(1-\mu^{2}),
\label{diffusion-coefficient}
\end{equation}
where $D_{0}$ is a constant indicating the magnetic turbulence
strength and $R$ is the particle rigidity. The constant $h$ is
needed to simulate the particles' ability to scatter through
$\mu=0$. In order to simulate the nonlinear effect to cause large
$D_{\mu\mu}$ at $\mu=0$, we use a relatively large value of $h=0.2$.
The constant $q$ is related to the power spectrum of the magnetic
field turbulence in the inertial range, chosen to be $5/3$ in the
model. Additionally, we assume that the two perpendicular diffusion
coefficients, $\lambda_x$ and $\lambda_y$, are the same and
independent of $\mu$. According to the observational determinations
and theoretical investigations of the mean free paths in previous
works \citep[see][and references therein]{He2013b}, we typically use
the radial mean free path $\lambda_{r}=0.3$ AU (corresponding to the
parallel mean free path $\lambda_{\parallel}=0.67$ AU at 1 AU) and
the perpendicular mean free paths $\lambda_{x}=\lambda_{y}=0.006$
AU. Actually, altering the values of the parameters in the model
will not qualitatively change the simulation results, i.e., the
longitudinally asymmetric distribution of the SEP fluxes, which will
be discussed later.

In order to numerically solve the focused transport equation
(\ref{transport-equation}), we utilize the time-backward Markov
stochastic process method \citep[e.g.,][]{Zhang1999}. This approach
could deal conveniently with an expanded source energy spectrum.
When using the time-backward stochastic process method, we trace
SEPs back to the initial time, and only those particles in the
source region at the initial time contribute to the statistics. The
five time-backward stochastic differential equations transformed
from the focused transport equation (\ref{transport-equation}) can
be described as follows:
\begin{eqnarray}
dX &=& \sqrt{2\kappa_{xx}}dW_{x}(s)-V_{x}^{sw}ds  \nonumber\\
dY &=& \sqrt{2\kappa_{yy}}dW_{y}(s)-V_{y}^{sw}ds  \nonumber\\
dZ &=& -(\mu V+V_{z}^{sw})ds  \nonumber\\
d\mu &=& \sqrt{2D_{\mu\mu}}dW_{\mu}(s)  \nonumber\\
{}&& -\frac{1-\mu^{2}}{2}\left[\frac{V}{L}+\mu\left(\frac{\partial
V_{x}^{sw}}{\partial x}+\frac{\partial V_{y}^{sw}}{\partial
y}-2\frac{\partial V_{z}^{sw}}{\partial z}\right)\right]ds  \nonumber\\
{}&& +\left(\frac{\partial D_{\mu\mu}}{\partial
\mu}+\frac{2D_{\mu\mu}}{M+\mu}\right)ds  \nonumber\\
dP &=& P\left[\frac{1-\mu^{2}}{2}\left(\frac{\partial
V_{x}^{sw}}{\partial x}+\frac{\partial V_{y}^{sw}}{\partial
y}\right)+\mu^{2}\frac{\partial V_{z}^{sw}}{\partial z}\right]ds,
\label{eq:stochastic-process}
\end{eqnarray}
where $(X,Y,Z)$ is the pseudo-position, $V$ is the pseudo-velocity,
$P$ is the pseudo-momentum, and $W_{x}(t)$, $W_{y}(t)$, and
$W_{\mu}(t)$ are Wiener processes. In our simulations, the Parker
interplanetary magnetic field $\textbf{B}$ is set so that its
magnitude at $1 AU$ is $5 nT$, and the solar wind speed is typically
set as $V^{sw}=400~km~s^{-1}$.

The source term $Q$ in Equation (\ref{transport-equation}) with a
power-law spectrum $\gamma$ is assumed as \citep{Reid1964}
\begin{equation}
Q(z<0.05AU,E_{k},\theta,\phi,t)=\frac{C}{t}\frac{E_{k}^{-\gamma}}{p^{2}}
\exp\left(-\frac{\tau_{c}}{t}-\frac{t}{\tau_{L}}\right)\xi(\theta,\phi),
\label{source}
\end{equation}
where $\xi(\theta,\phi)$ is a function of the heliographic latitude
$\theta$ and longitude $\phi$ that describes the spatial
distribution of SEP source strength, $E_k$ is the kinetic energy of
the source particles, and $\tau_{c}$ and $\tau_{L}$ denote the rise
and decay timescales of source release profile, respectively.
Although the injection form in Equation (\ref{source}) is originally
based on an assumption of lateral diffusion in the corona, it can be
used as a short pulse of source particle injection, especially when
the shock is near the Sun. Typically, the values of $\tau_{c}$ and
$\tau_{L}$ are a few hours, which are much shorter than the
timescale to reach the maximum flux or the decay phase of SEP events
\citep{Zhang2009}. Therefore, $\tau_{c}$ and $\tau_{L}$ only affect
the flux profiles at the beginning, but not the long-term when
$t\gg\tau_{L}$. In this work, we typically set $\gamma=3$,
$\tau_{c}=0.1$ days and $\tau_{L}=0.25$ days. As we know, typical
sizes of SEP sources (flares or CMEs) are tens of degrees wide in
heliographic latitude and longitude
\citep[e.g.,][]{Hundhausen1993,Maia2001,Wang2006}. It is not easy to
find out the exact sizes of SEP sources; in the simulations, we set
SEP sources at $r=0.05 AU$ with limited coverage of $30^{\circ}$ in
latitude and longitude. Furthermore, we assume a uniform spatial
distribution in the source regions. In addition, an outer absorptive
boundary of pseudoparticles is set at $r=50 AU$.

To obtain the SEP flux for each case, we simulate $6\times 10^7$
pseudoparticles on a super-computer cluster with message passing
interface. Generally, the unit of omnidirectional flux is used as
$cm^{-2}-sr^{-1}-s^{-1}-MeV^{-1}$; in this work, however, we use an
arbitrary unit for convenience in plotting figures.

\section{Numerical Results}
For each particle energy investigated, we simulate 18 cases
corresponding to different source locations to show the effect of
solar source location on the SEP flux observed in the interplanetary
space. All the solar particle sources have the same coverage, i.e.,
$30^{\circ}$ in latitude and longitude, but their centers are
located at various heliographic longitudes in the solar equator or
at $90^{\circ}$ colatitude. Specifically, the longitude separations
between the centers of the 18 solar sources and the magnetic
footpoint of the observer are set to be the following values:
$0^{\circ}$, $20^{\circ}$ west, $20^{\circ}$ east, $40^{\circ}$
west, $40^{\circ}$ east, $\ldots$, $160^{\circ}$ west, $160^{\circ}$
east, $180^{\circ}$ east. All the other conditions and parameters,
such as the source particle intensity, solar wind speed, turbulent
magnetic field, etc., are the same for each case. In the
simulations, we typically investigate $20$ MeV, $50$ MeV, and $100$
MeV protons detected at 1 AU, $90^{\circ}$ colatitude.

\begin{figure}
 \epsscale{1.0}
 \plotone{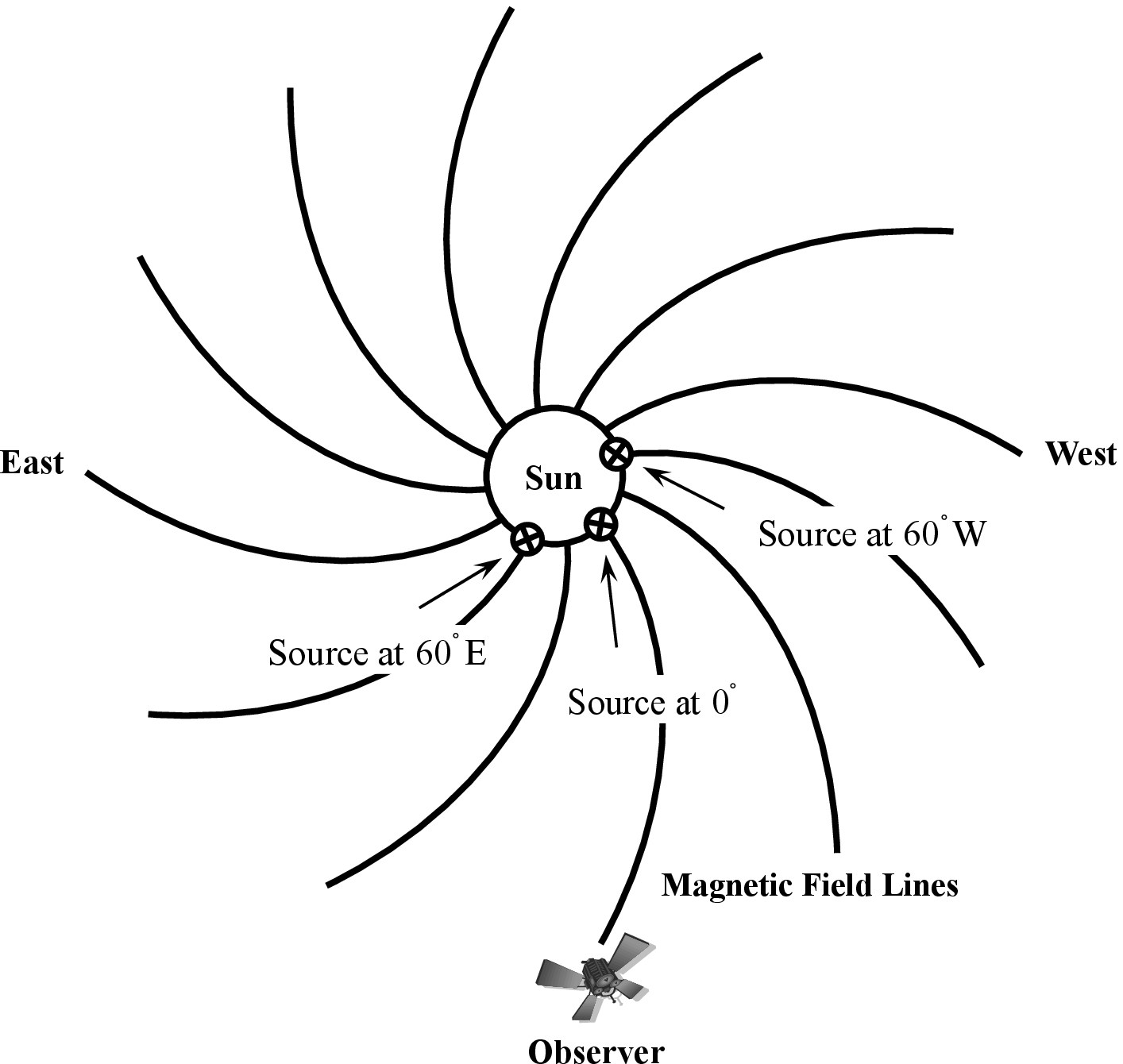}
 \caption{Diagram to illustrate the longitudinal locations of
the particle sources in the solar equator relative to the magnetic
footpoint of the observer, which is also located at $90^{\circ}$
colatitude. The center of the middle solar source is connected
directly to the spacecraft by the interplanetary magnetic field
line, whereas the centers of the solar sources on the left and right
are $60^{\circ}$ east and $60^{\circ}$ west, respectively, away from
the magnetic footpoint of the observer. \label{diagram}}
\end{figure}

A diagram, i.e., Figure \ref{diagram}, serving as an example, is
presented to illustrate the longitudinal locations of the particle
sources on the solar surface relative to the magnetic footpoint of
the observer, which is located at $90^{\circ}$ colatitude. In the
scenario of the diagram, we can see that the center of the middle
source on the Sun is connected directly to the spacecraft by
interplanetary magnetic field line, whereas the centers of the solar
sources on the left and right are $60^{\circ}$ east and $60^{\circ}$
west, respectively, away from the magnetic footpoint of the
observer.

\begin{figure}
 \epsscale{1.0}
 \plotone{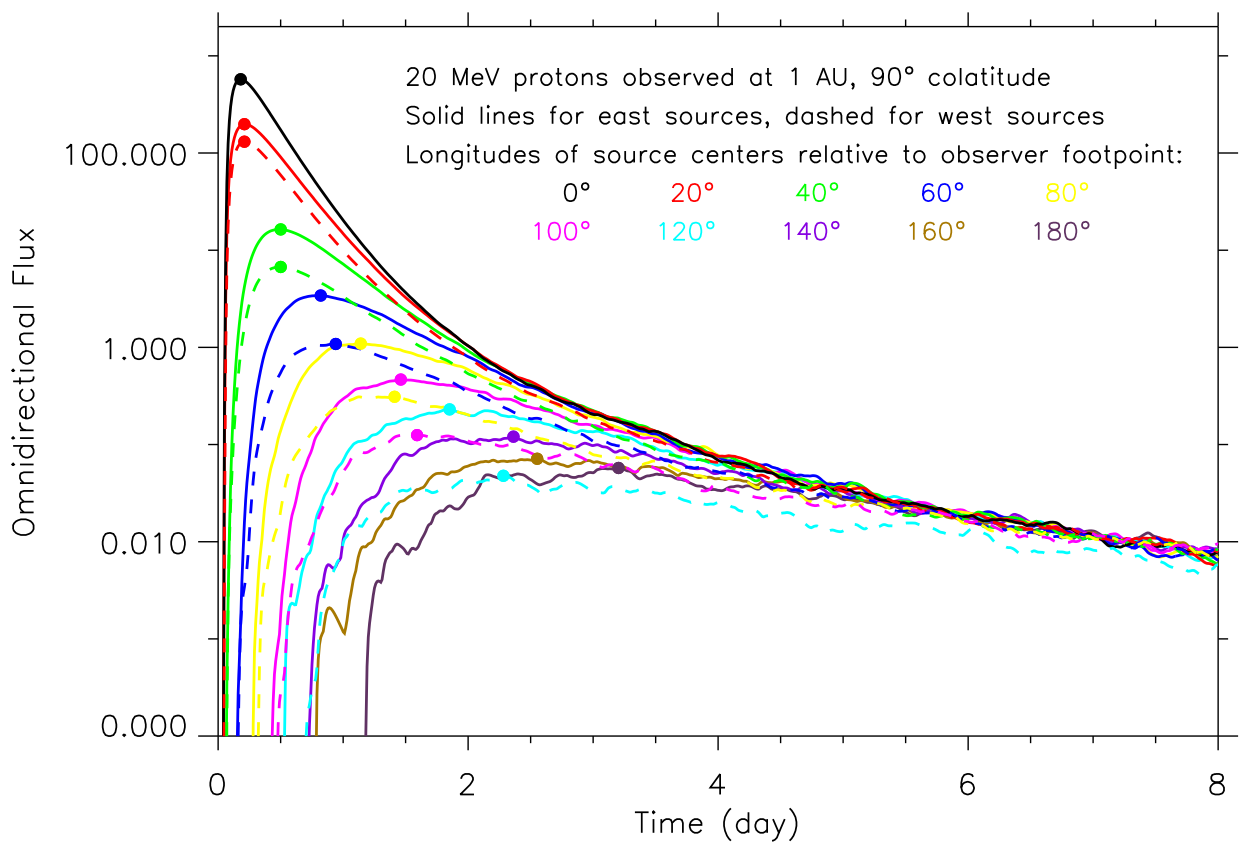}
 \caption{Flux-time profiles of the $20$ MeV SEPs originating from the
different solar sources with different longitudinal locations
relative to the magnetic footpoint of the observer at 1 AU,
$90^{\circ}$ colatitude. With the same longitude separations between
the magnetic field footpoint of the observer and solar sources, the
SEP fluxes from solar sources located on the eastern side of the
observer footpoint are larger than those from solar sources located
on the western side. In addition, the SEP reservoir phenomenon is
reproduced. \label{flux-20MeV}}
\end{figure}

Figure \ref{flux-20MeV} shows the flux-time profiles of the $20$ MeV
SEPs originating from the different solar sources with different
longitudinal locations relative to the magnetic footpoint of the
observer at 1 AU, $90^{\circ}$ colatitude. In Figure
\ref{flux-20MeV}, the solid and dashed lines denote the SEP
omnidirectional fluxes originating from the east and west solar
sources, respectively. The different colors of the curves represent
different longitudinal distances between the magnetic footpoint of
the observer and the solar sources. The $0^{\circ}$ relatively
longitudinal distance denotes the case where the nominal magnetic
footpoint of the observer and the solar source center locate at the
same heliographic longitude. As we can see, the SEP flux in this
case is the largest among all the cases investigated. This means
that when the observer in the interplanetary space is nearly
directly connected to the solar source by heliospheric magnetic
field lines, the SEP flux is larger than that otherwise. However, a
careful investigation based on our numerical simulations indicates
that the maximum SEP flux shifts from the location of best magnetic
connection and toward the east SEP sources with centers on relative
longitudes $\sim-0.7^{\circ}-\sim-1.5^{\circ}$. We note that the
shift extents depend on the ratio
$\lambda_{\perp}/\lambda_{\parallel}$ of the perpendicular to the
parallel mean free paths, the geometry of the magnetic spiral, the
energy of SEPs, the coverage of the SEP sources, and the particle
spatial distribution in the sources. In Figure \ref{flux-20MeV}, we
can also see that the farther the magnetic field footpoint of the
observer is away from the solar source, the smaller is the SEP flux
observed and also the later the SEP event arrives at the observer's
position \citep[see also][]{He2011b}. Moreover, with the same
longitude separations between the magnetic field footpoint of the
observer and solar sources, the SEP fluxes from solar sources
located on the eastern side of the observer footpoint are larger
than those from solar sources located on the western side, and the
times of onsets and maximum fluxes of SEP events from east sources
are earlier than those from west sources. In other words, there
exists an east-west longitudinally asymmetric distribution of SEPs
in the heliosphere. In addition, we can obviously observe that in
the decay phase the flux profiles of all the SEP cases have almost
the same decay rate, which is called the SEP reservoir phenomenon in
spacecraft observations. Therefore, our numerical simulations with a
series of different SEP cases almost covering $360^{\circ}$ in
heliographic longitude successfully reproduce the so-called SEP
reservoir. The SEP reservoirs are observed by spacecraft at both low
and high heliolatitudes, and by spacecraft at very different
heliolongitudes and radial distances. In our simulations with
perpendicular diffusion, we reproduce various SEP reservoirs (some
results not shown here) at different heliospheric locations
(longitude, latitude, and radial distance) without invoking the
hypothesis of a reflecting boundary or diffusion barrier. Actually,
it is difficult to imagine such an ``overwhelming" reflecting
boundary or diffusion barrier covering all the longitudes, all the
latitudes, and even all the radial distances. Therefore, the
perpendicular diffusion should be the dominant, if not the only,
physical mechanism responsible for the formation of the SEP
reservoirs.

\begin{figure}
 \epsscale{1.0}
 \plotone{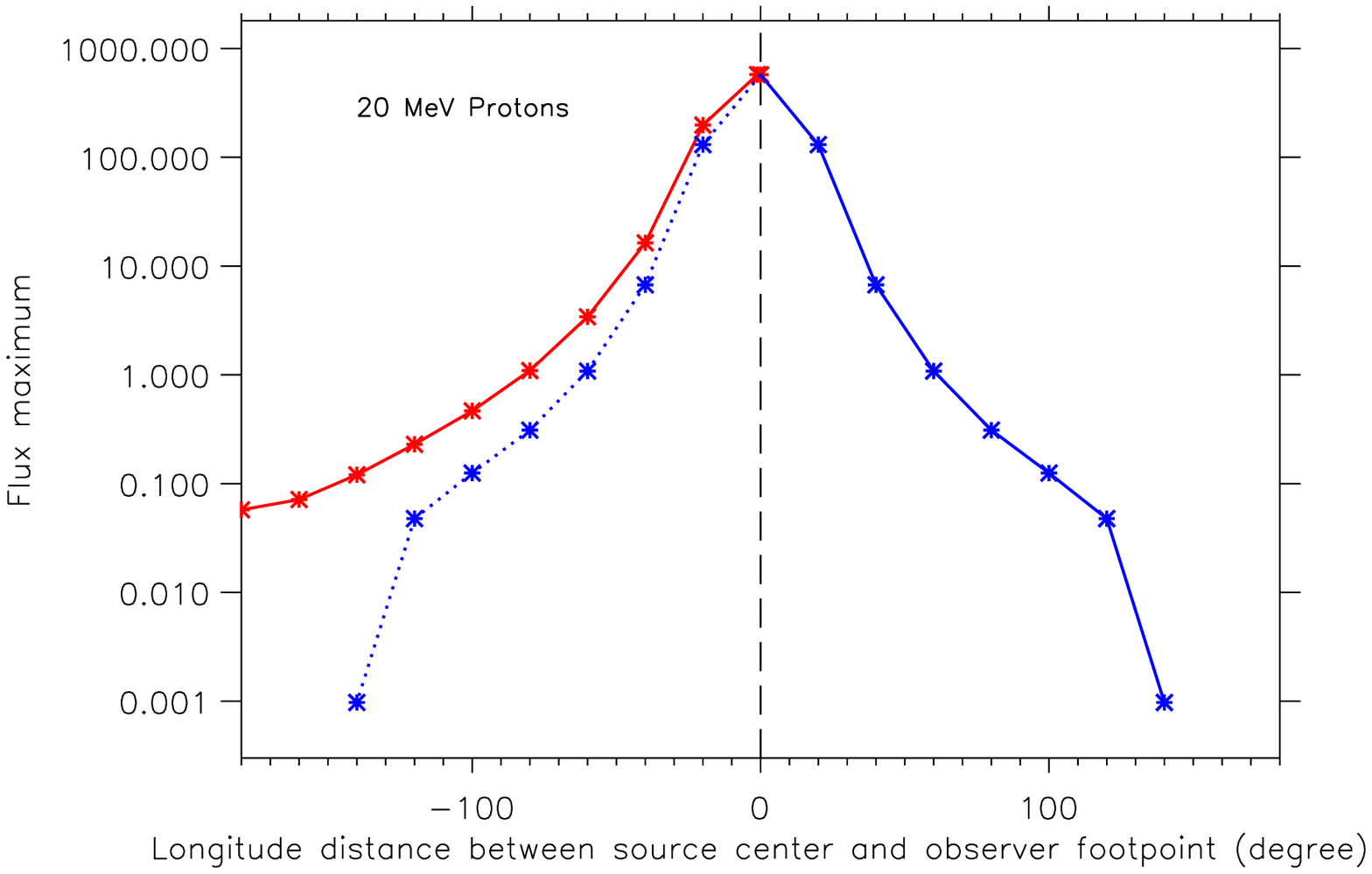}
 \caption{Flux maxima (peak fluxes) of the $20$ MeV SEPs originating from the different
 solar sources with different longitudinal locations relative to the magnetic
 footpoint of the observer at 1 AU, $90^{\circ}$ colatitude. With the
same longitude separations between the solar source centers and the
magnetic footpoint of the observer, the flux maxima of the SEP
events originating from solar sources located on the eastern (left)
side of the observer footpoint are systematically larger than those
of the SEP events originating from sources located on the western
(right) side. \label{ew-20MeV}}
\end{figure}

The circles with different colors on the flux profiles in Figure
\ref{flux-20MeV} denote the peak fluxes of the corresponding SEP
cases. We extract the information of the peak fluxes and the
relevant longitudinal distances in the SEP cases investigated from
Figure \ref{flux-20MeV}, and present it separately. In Figure
\ref{ew-20MeV}, we show the peak fluxes of the $20$ MeV SEPs
originating from the different solar sources with different
longitudinal locations relative to the magnetic footpoint of the
observer at 1 AU, $90^{\circ}$ colatitude. The X-axis in Figure
\ref{ew-20MeV} is the relatively longitudinal distance $\phi_{r}$
between the solar source center and the magnetic footpoint of the
observer. The vertical dashed line in the middle of Figure
\ref{ew-20MeV} denotes the $0^{\circ}$ relatively longitudinal
distance, where the nominal magnetic footpoint of the observer and
the solar source center locate at the same heliographic longitude.
As we can see, the peak flux in the SEP case with about $0^{\circ}$
relatively longitudinal distance is the largest among the cases
investigated. This means that when the observer in the
interplanetary space is nearly directly connected to the solar
source by heliospheric magnetic field lines, the SEP peak flux is
larger than that otherwise. However, a careful investigation
indicates that the maximum SEP peak flux shifts from the place of
best connection and toward the east SEP sources with centers on
relative longitudes $\sim-0.7^{\circ}-\sim-1.5^{\circ}$. We connect
the peak fluxes of SEPs in the simulated cases on the left-hand and
right-hand halves with the red and blue curves, respectively. It can
obviously be seen from Figure \ref{ew-20MeV} that the farther the
solar source center is away from the magnetic footpoint of the
observer, the smaller is the peak flux of SEPs observed. We further
mirror the peak fluxes of SEPs on the right-hand half to the
left-hand half with a blue dotted curve. We can obviously observe
that with the same longitude separations between the solar source
centers and the magnetic footpoint of the observer, the peak fluxes
of SEP events originating from solar sources located on the eastern
side of the observer footpoint are systematically larger than those
of the SEP events originating from sources located on the western
side. Therefore, our relatively complete model calculation of SEP
propagation with the effect of perpendicular diffusion shows that
there exists an east-west asymmetry in the SEP intensities.

\begin{figure}
 \epsscale{1.0}
 \plotone{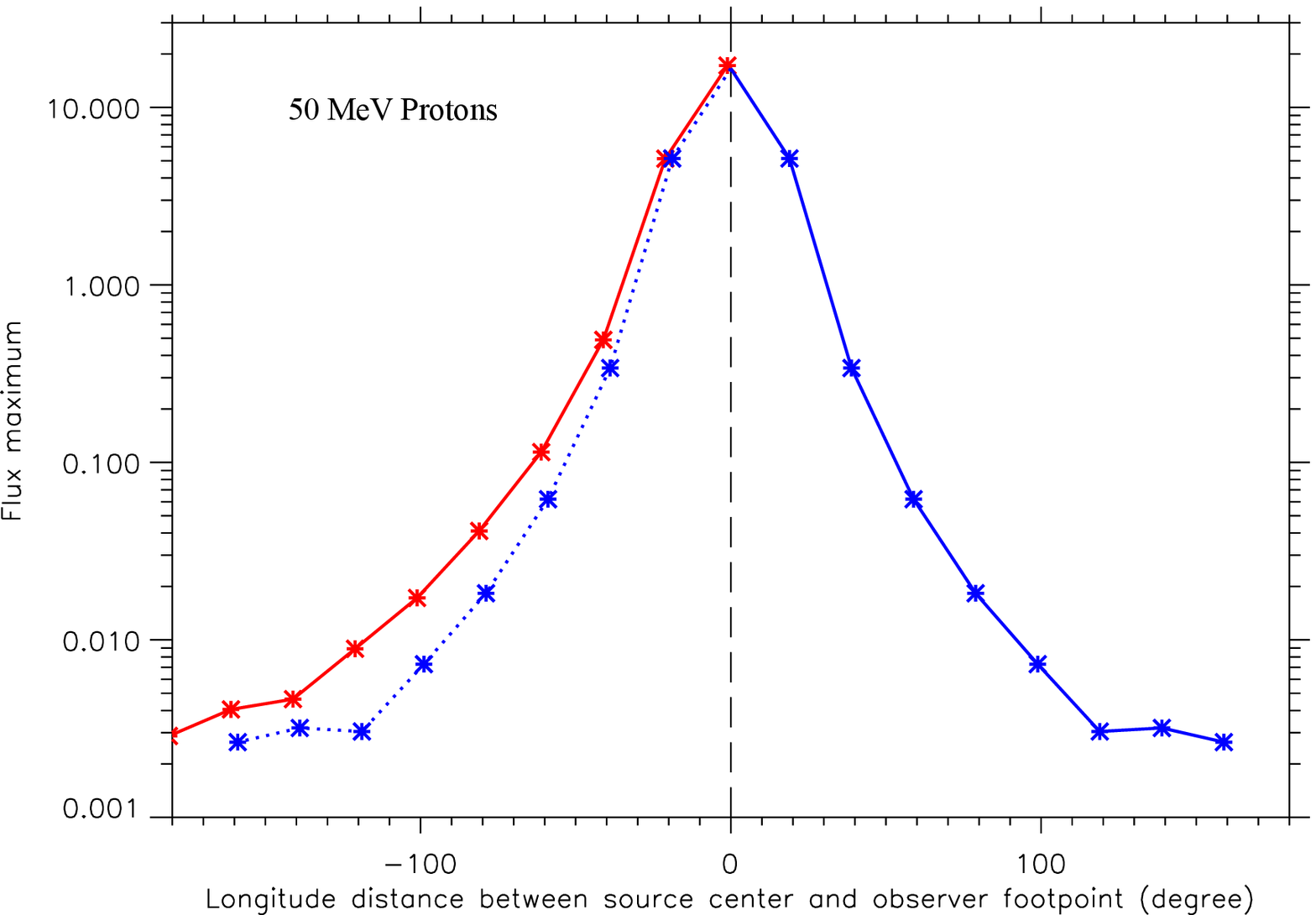}
 \caption{Flux maxima (peak fluxes) of the $50$ MeV SEPs originating from the different
 solar sources with different longitudinal locations relative to the magnetic
 footpoint of the observer at 1 AU, $90^{\circ}$ colatitude. With the
same longitude separations between the solar source centers and the
magnetic footpoint of the observer, the flux maxima of the SEP
events originating from solar sources located on the eastern (left)
side of the observer footpoint are systematically larger than those
of the SEP events originating from sources located on the western
(right) side. \label{ew-50MeV}}
\end{figure}

\begin{figure}
 \epsscale{1.0}
 \plotone{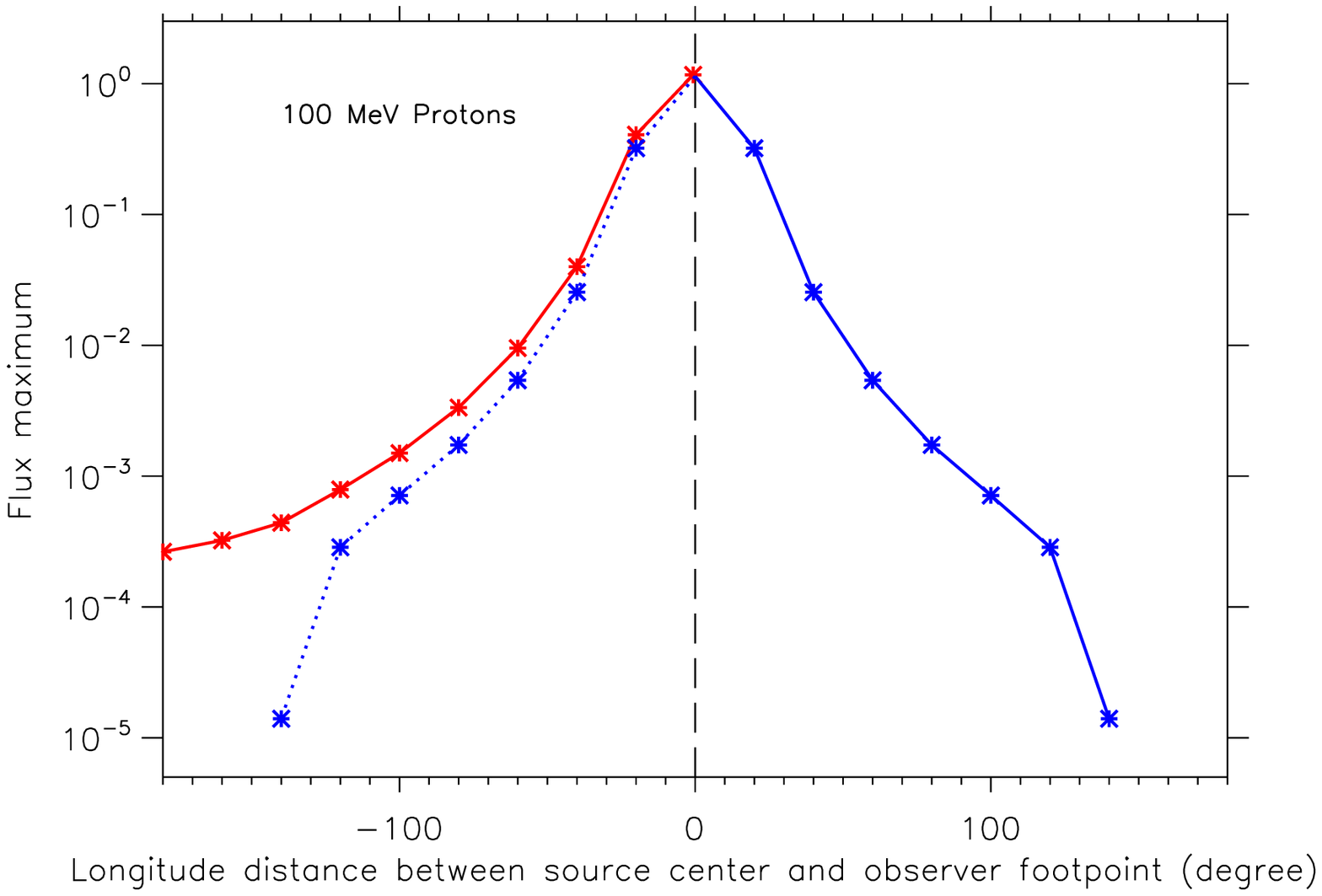}
 \caption{Flux maxima (peak fluxes) of the $100$ MeV SEPs originating from the different
 solar sources with different longitudinal locations relative to the magnetic
 footpoint of the observer at 1 AU, $90^{\circ}$ colatitude. With the
same longitude separations between the solar source centers and the
magnetic footpoint of the observer, the flux maxima of the SEP
events originating from solar sources located on the eastern (left)
side of the observer footpoint are systematically larger than those
of the SEP events originating from sources located on the western
(right) side. \label{ew-100MeV}}
\end{figure}

Figure \ref{ew-50MeV} shows the peak fluxes of the $50$ MeV SEPs
originating from the different solar sources with different
longitudinal distances relative to the magnetic footpoint of the
observer at 1 AU, $90^{\circ}$ colatitude. As in Figure
\ref{ew-20MeV}, we can also clearly observe that with the same
longitudinal distances between the solar source centers and the
magnetic footpoint of the observer, the peak fluxes of the SEP
events originating from solar sources located on the eastern side of
the observer footpoint are systematically larger than those of the
SEP events originating from sources located on the western side.
Therefore, the numerical simulation of the transport process of $50$
MeV SEPs also shows that there exists an east-west longitudinally
asymmetric distribution of SEPs in the heliosphere. We note that the
subtle difference between Figure \ref{ew-20MeV} and Figure
\ref{ew-50MeV} results from the interplay of diffusion coefficients
$\kappa_{\perp}$ and $\kappa_{\parallel}$ as well as the weak
statistics of SEP fluxes in the cases with long longitudinal
distances between the solar sources and the magnetic footpoint of
the observer. Figure \ref{ew-100MeV} shows the peak fluxes of the
$100$ MeV SEPs originating from the different solar sources with
different longitudinal distances relative to the magnetic footpoint
of the observer at 1 AU, $90^{\circ}$ colatitude. It can also
evidently be seen that with the same longitudinal distances between
the solar source centers and the magnetic footpoint of the observer,
the peak fluxes of the SEP events originating from solar sources
located on the eastern side of the observer footpoint are
systematically larger than those of the SEP events originating from
sources located on the western side.

\section{Implications of Simulation Results}
Generally, an SEP event with a higher peak flux will reveal a larger
intensity relative to other events through the entire evolution
process (see Figure \ref{flux-20MeV} and the results in
\citet{He2011b}). In general, the probability of the SEP events
being observed by spacecraft near Earth's orbit depends on the
intensities of the particles in the events. An intense SEP event
with high flux will experience relatively weak influences from
interplanetary structures (such as magnetic cloud, heliospheric
current sheet, corotating interaction region, etc.) during the
propagation process in the heliosphere. However, the not so intense
SEP events with small fluxes will be significantly affected by the
interplanetary structures and then to weaken or even dissipate
before entering Earth's orbit. For example, numerous observations
have shown that a magnetic cloud can considerably affect the
propagation conditions of SEPs and low-energy cosmic rays, which is
commonly observed as a decrease in the fluxes of the energetic
particles \citep[e.g.,][]{Blanco2013}. Consequently, intense SEP
events would have a larger probability of reaching Earth and being
observed by spacecraft at 1 AU; whereas the relatively weak SEP
events would have a lower probability to arrive at Earth and to be
recorded by spacecraft.

As shown in the simulation results above, the SEP events originating
from solar sources located on the eastern side of the magnetic
footpoint of the observer reveal larger fluxes than those
originating from solar sources located on the western side.
Therefore, the SEP events originating from solar sources located at
eastern relative longitudes would have a higher probability to reach
Earth and to be observed by spacecraft near 1 AU. On the contrary,
the SEP events from solar sources on the west side relative to the
magnetic footpoint of observer will be more easily influenced by the
interplanetary medium and structures, and then to weaken or even
dissipate during their propagation processes. As a result, the
spacecraft at 1 AU would have a larger probability to miss the SEP
events originating from solar sources located at western relative
longitudes.

\section{Discussion}
\begin{figure}
 \epsscale{1.0}
 \plotone{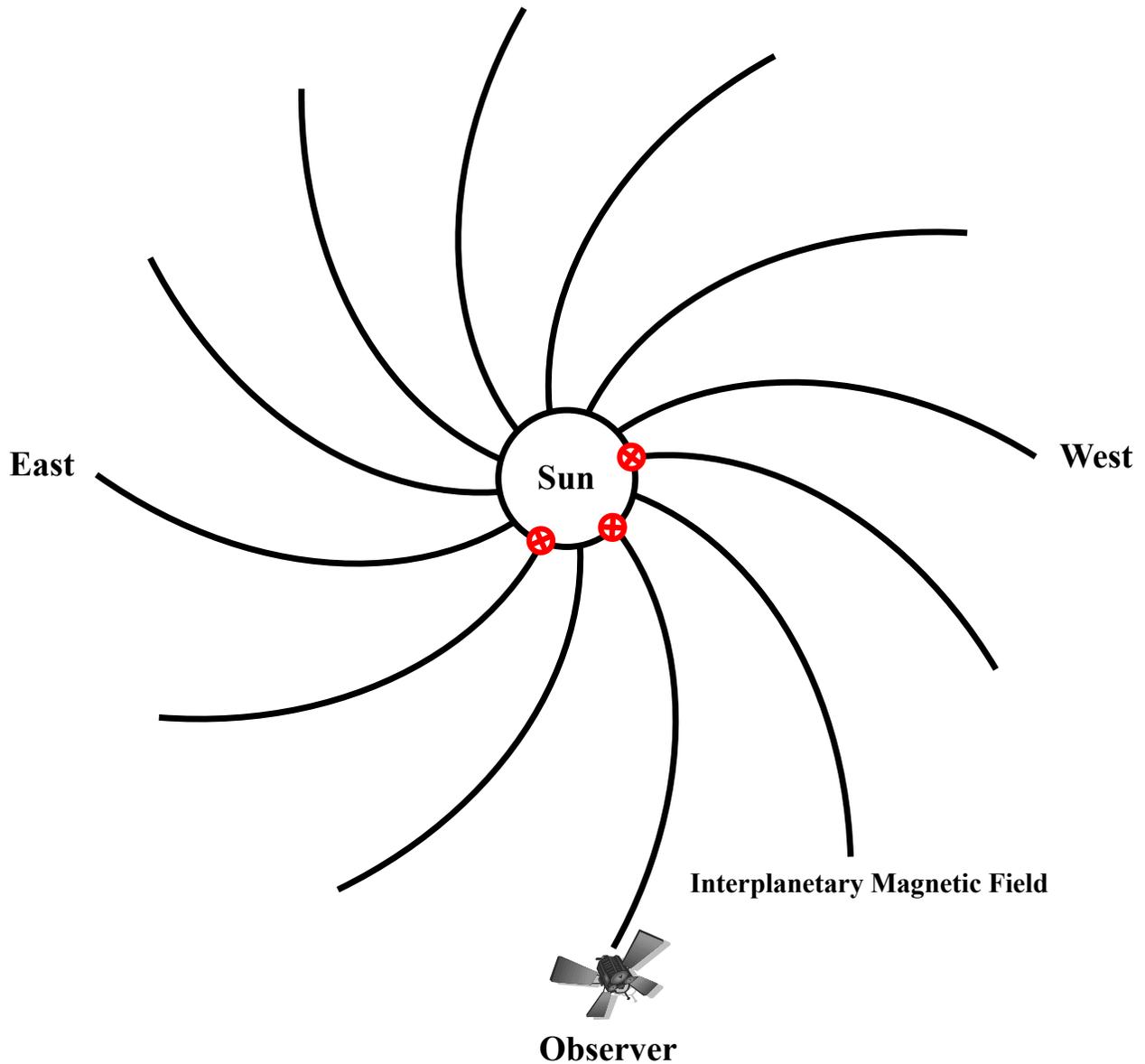}
 \caption{Sketch to show the east-west azimuthal asymmetry in the topology of
the Parker interplanetary magnetic field. Due to the effects of the
perpendicular diffusion and the azimuthally asymmetric geometry of
the interplanetary magnetic field, with the same longitudinal
distances between the solar sources and the magnetic footpoint of
the observer, the SEPs originating from sources located on the
eastern side of the observer footpoint are easier and more frequent
than those from sources located on the western side to arrive at the
observer in the interplanetary space. \label{Parker-spiral}}
\end{figure}

Generally, as viewed from above the north pole of the Sun, the
Parker spiral magnetic field lines originating from the solar
surface would meander clockwise to somewhere in the interplanetary
space with more eastern longitudes than their footpoints on the
solar surface. This scenario can be seen in Figure
\ref{Parker-spiral}. As we can see, there is an east-west azimuthal
asymmetry in the topology of the Parker interplanetary magnetic
field. According to the theoretical and observational investigations
\citep[e.g.,][]{Matthaeus2003,Bieber2004,He2012b}, for physical
conditions representative of the solar wind, the perpendicular
diffusion coefficient would be a few percent of the parallel
diffusion coefficient in magnitude. In general, SEPs would mainly
transport along the average interplanetary magnetic field after they
were produced from the surface of the Sun or a CME-driven shock.
Therefore, when the observer is nearly directly connected to the
solar source by the interplanetary magnetic field lines, the
omnidirectional flux and peak flux of SEPs are larger than those in
the cases where the observer is not connected directly to the solar
sources. In general, the farther the solar source center is away
from the magnetic footpoint of the observer, the smaller are the
omnidirectional flux and peak flux of SEPs observed. In addition to
the parallel diffusion, the SEPs will experience perpendicular
diffusion to cross the heliospheric magnetic field lines during
their transport processes in the interplanetary space. This is why
the SEP events can be observed by a spacecraft that is not directly
connected to the acceleration regions. In the limit of a point
source release of SEPs and no perpendicular diffusion, the particles
would also transport in longitude as a result of the field line
rotating with the Sun. However, as \citet{Giacalone2012} pointed
out, in this case, the intensity profile of the event observed near
1 AU would be a spike of SEPs only seen when the field line
containing the particles crosses the observer and would be zero for
all other times, which is inconsistent with the usually observed
shape of intensity profile of SEPs. A careful investigation based on
our numerical simulations indicates that the maximum SEP flux shifts
from the location of best magnetic connection and toward the east
SEP sources with centers on relative longitudes
$\sim-0.7^{\circ}-\sim-1.5^{\circ}$. Note that the shift extents
depend on the ratio $\lambda_{\perp}/\lambda_{\parallel}$, the
geometry of the magnetic spiral, the energy of SEPs, the coverage of
the SEP sources, and the particle spatial distribution in the
sources.

Due to the effects of the perpendicular diffusion and the
azimuthally asymmetric geometry of the heliospheric magnetic field,
with the same longitudinal distances between the solar sources and
the magnetic footpoint of the observer, the SEPs originating from
sources located on the eastern side of the observer footpoint are
easier and more frequent than those from sources located on the
western side to arrive at the observer in the interplanetary space.
Consequently, with the same longitudinal distances between the solar
sources and the observer footpoint, the fluxes of the SEP events
associated with the solar sources located east are systematically
larger than those of the SEP events associated with the sources
located west. Accordingly, the peak intensities of the SEP events
originating from solar sources located on the eastern side of the
observer footpoint are also larger than those of the SEP events
originating from sources located on the western side. Therefore, we
propose that the longitudinally asymmetric distribution of SEPs
results from the east-west azimuthal asymmetry in the topology of
the Parker interplanetary magnetic field as well as the effects of
perpendicular diffusion on the transport processes of SEPs in the
heliosphere. Furthermore, the interplay between the perpendicular
diffusion ($\lambda_{\perp}$) and the parallel diffusion
($\lambda_{\parallel}$) plays a very important role in determining
the azimuthal asymmetry extents of the SEP distribution in the
interplanetary space.

\section{Summary and Conclusions}
In the previous work of \citet{He2011b}, the authors reported and
predicted that with the same longitude separation between the
magnetic footpoint of observer and the solar sources, the SEPs
produced and released from the sources located east are detected
earlier with larger fluxes than those associated with the sources
located west. This interesting phenomenon was called the ``east-west
azimuthal asymmetry of SEPs" by \citet{He2011b}. The figures in
\citet{Giacalone2012} using Parker's equation also displayed
east-west asymmetry in the SEP distribution. In this work, we
systematically study the longitudinally asymmetric distribution of
SEPs in view of numerical investigations. We carry out a series of
relatively complete model calculations of SEP propagation with
perpendicular diffusion in a three-dimensional interplanetary
magnetic field. Our simulation results show that with the same
longitude separation between the solar sources and the magnetic
footpoint of the observer, the flux of SEPs released from the solar
source located east is systematically higher than that of SEPs
originating from the source located west. We conclude that the
longitudinally asymmetric distribution of SEPs results from the
east-west azimuthal asymmetry in the topology of the Parker
interplanetary magnetic field as well as the effects of
perpendicular diffusion on the transport processes of SEPs in the
heliosphere. The interplay between the perpendicular diffusion
($\lambda_{\perp}$) and the parallel diffusion
($\lambda_{\parallel}$) plays an important role in determining the
extents of the azimuthal asymmetry of the SEP distribution in the
heliosphere.

In general, the probabilities of the SEP events being observed by
spacecraft in the interplanetary space are proportional to the
particle flux intensities in the events. Consequently, the SEP
events originating from solar sources located at the eastern
relative longitudes would be detected and recorded with somewhat
higher probabilities by the observer near 1 AU. Therefore, the
results presented in this work will be valuable for understanding
the solar-terrestrial relations and useful in space weather
forecasting.


\acknowledgments

We thank the anonymous referee for valuable comments. This work was
supported in part by the National Natural Science Foundation of
China under grants 41204130, 41474154, 41321003, and 41131066, the
National Important Basic Research Project under grant 2011CB811405,
the Chinese Academy of Sciences under grant KZZD-EW-01-2, the China
Postdoctoral Science Foundation under grants 2011M500381 and
2012T50131, the Open Research Program from Key Laboratory of
Geospace Environment, Chinese Academy of Sciences, and the Key
Laboratory of Random Complex Structures and Data Science, CAS. H.-Q.
He gratefully acknowledges the support of K.C.Wong Education
Foundation.


\clearpage


\clearpage

\clearpage

\clearpage

\clearpage

\clearpage

\clearpage


\end{document}